\documentclass[pra,aps,twocolumn,showpacs,epsfig]{revtex4}
\usepackage{color}
\usepackage{graphicx,amssymb,epsfig,epsf,amsmath}
 
\begin{document}
\title{Level spacing statistics and spectral correlations in the diffuse van der Waals clusters}

\author{
S. K. Haldar$^{1}$, B. Chakrabarti$^{2}$\footnote{Present address: Department of Physics, Presidency University, 86/1 College Street, Kolkata 700073}, 
N. D. Chavda$^{3}$, T. K. Das$^{4}$, S. Canuto$^{5}$, V. K. B. Kota$^{6}$ 
}

\affiliation{
$^{1}$Department of Physics, Lady Brabourne College, P-$\frac{1}{2}$ Suhrawardi Avenue, Kolkata-700017, India.\\
$^{2}$Department of Physics, Kalyani University, Kalyani, Nadia 741235, West Bengal, India.\\
$^{3}$Applied Physics Department, Faculty of technology and Engineering, Maharaja Sayajirao University of Baroda, Vadodara 390 001, India.\\
$^{4}$Department of Physics, Calcutta University, 92 A. P. C. Road, Kolkata-700009, India.\\
$^{5}$Instituto de Fisica, Universidade of S\~ao Paulo, CP 66318, 05315-970, S\~ao Paulo, Brazil.\\
$^{6}$Physical Research Laboratory, Navarangpura, Ahmedabad 380009, India.
}

\begin{abstract}
We present a statistical analysis of eigenenergies and discuss several measures of spectral fluctuations and spectral correlations for 
the van der Waals clusters of different sizes. We show that the clusters become
more and more complex with increase in cluster size. We study nearest-neighbour level spacing distribution $P(s)$, 
the level number variance $\Sigma^2(L)$, and the Dyson-Mehta $\Delta_3-$statistics for various cluster sizes.
For large clusters we find that although the Bohigas-Giannoni-Schmit (BGS) conjecture seems to be valid, it does not exhibit true signatures
of quantum chaos. However contrasting conjecture of Berry and Tabor is observed with 
smaller cluster size. For small number of bosons, we observe the existence of large number of quasi-degenerate states in low-lying excitation which exhibits the 
Shnirelman peak in $P(s)$ distribution. We also find a narrow region of intermediate spectrum which can be described by semi-Poisson statistics whereas the 
higher levels are regular and exhibit Poisson statistics. These observations are further 
supported by the analysis of the distribution of the ratio of consecutive level spacings $P(r)$ which is independent of unfolding procedure and thereby 
provides a tool for more transparent comparison with experimental findings than $P(s)$. Thus our detail numerical study clearly shows 
that the van der Waals clusters become more correlated with the increase in cluster size.
\end{abstract}
\maketitle
\section{Introduction}

 Weakly bound few-body systems are being studied since a long time back and have achieved revived interest recently as the physics of such 
weakly bound systems can be investigated experimentally in ultracold atomic gases~\cite{Kramer}. Utilizing the Feshbach resonance, the effective 
inter-atomic interaction can be changed essentially to any desired values~\cite{Donley,Chin}. The recent experiments on cold atoms also provide 
evidence of the existence of large weakly bound clusters. Thus our present study is motivated by the recent experiments on ultracold Bose gas. 
We treat the three-dimensional bosonic cluster with maximum up to $N = 40$ Rb atoms interacting through two-body van der Waals potential. 
Alkali atoms, specially Rb atoms, are good candidates for laser manipulation and to observe Bose-Einstein condensate~\cite{Anderson}. 
At ultracold temperature the interatomic interaction is fairly well represented by a single parameter $a_s$, the $s-$wave scattering length. 
For our present system we keep $a_s = 100$ $a_0$ which corresponds to the JILA experiment~\cite{Anderson}. 
Thus the system is weakly interacting, and diffuse as the average size of the 
cluster increases with cluster size. The binding of such $N-$body cluster is provided by the two-body van der Waals potential having a short range 
repulsive core below a cutoff radius and a $\frac{-C_6}{r^6}$ tail which represents the long range attractive interaction.

The stability of such $N-$body clusters, their energetics and various structural properties are recently studied~\cite{Pankaj}. We propose the use 
of two-body basis function to describe various properties of bosonic clusters. With more than three particles the system becomes more complex as the 
number of degrees of freedom increases. We have investigated correlations between energies of the $N$ and $(N-1)$ systems and observe the generalized 
Tjon line~\cite{Pankaj} for large cluster. 
Now we consider the spectral statistics and spectral correlation of the atomic clusters of different sizes as 
these contain rich physics and also plays an pivotal role 
to establish 
the universal properties of quantum systems. Berry and Tabor conjectured that 
the fluctuation property of energy levels of a quantum system whose classical analog is regular, is characterised by Poisson statistics~\cite{Berry}. Whereas, 
the fluctuation property of energy levels of a quantum system whose corresponding classical dynamical system is fully chaotic  obeys the Bohigas-Giannoni-Schmit (BGS) 
conjecture~\cite{Bohigas}. This tells that Gaussian orthogonal ensemble (GOE) or Gaussian unitary ensemble (GUE) or Gaussian sympletic ensemble (GSE) 
statistics of random matrix theory, depending on time reversal symmetry and rotational symmetry of the system, will 
describe the fluctuation properties. However this conjecture is often interpreted in another way and the observation of level repulsion in 
the spectrum is treated as an indication of the non-integrability of the system. The Poisson distribution implies complete randomness in the relative 
positions of energy levels as they are completely uncorrelated. On the other hand Wigner distribution implies strong correlation among the energy levels. 

Earlier the spectral properties of many different quantum systems like atoms, atomic nuclei, quantum billiards have been 
studied~\cite{Casati, Haake, Seligman, Zimmermann, Tanner, Sakhr, Brody, Kota, Gomez}. Also some attempts have been made for non-interacting many-bosons 
and interacting bosonic system~\cite{Munoz, Leclair, Chavda, Vyas}. 
Recently we have reported the level spacing distribution of ultra-cold interacting bosons trapped in a harmonic potential~\cite{Barnali,Kamalika1,Kamalika2}. 
We found intriguing effect of both the 
interatomic interaction and the trap and observed deviation from the BGS cojecture. In this paper we are interested in similar type of calculation in 
the van der Waals bosonic clusters. Unlike the Bose-Einstein condensate where the external trapping provides the stability of the condensate, 
the van der Waals clusters are bound due to the van der Waals interaction. In the very dilute condition one may treat it as a uniform Bose gas. Apart from the 
experimental interest, this kind of systems are also challenging for the following reasons. First, solving the many-body Schr\"odinger equation itself is a 
challenging numerical task due to many degrees of freedom and the obvious question is what kind of approximation is to be valid for the description of such clusters. 
Secondly for large cluster size when the system becomes very much correlated, one may expect Wigner type spectral distribution. However it needs an
exhaustive study as level repulsion in the energy spectrum may not always lead to Wigner distribution which signifies chaos. It indicates that one may need 
to use some deformed GOE type of distribution for the correct description of nonintegrable but non-chaotic system.
We propose to study several measures of spectral fluctuations and spectral correlation to determine the degree of influence of the interatomic interaction.
This kind of study is also relevant as
the statistical fluctuation can be directly observed experimentally in the context of ultracold Bose gases. 
We calculate nearest neighbour level spacing distribution (NNSD) $P(s)$, the level number variance $\Sigma^2(L)$ 
and the Dyson-Mehta $\Delta_3$-statistics~\cite{Mehta} for 
various cluster sizes. However all these measures require unfolding of the spectrum to remove variation in the density of energy levels in different parts of 
the spectrum. We can either unfold the spectrum of each member of the ensemble 
separately and form ensemble averaged NNSD or a single unfolding function can be used for all the members of the ensemble. Depending on the unfolding procedure, 
the final outcome of NNSD may vary. Moreover suitable unfolding function is not always known a priori and generally is approximated by higher order polynomials. 
Therefore to verify the outcome of the NNSD, we further analyze the distribution of quotients of successive spacings $P(r)$ which does not require any unfolding and 
is independent of the energy level density.

The paper is organised as follows. In Section II, we introduce the 
many-body potential harmonic expansion method.
Section III discusses the numerical results and Section IV concludes with 
the summary of our work.

\section{Methodology:Many-body calculation with potential 
harmonic basis}

To study the spectral statistics and different spectral correlations we need to calculate a large number of energy levels of the diffuse Rb cluster. We
approximately solve the full many-body Schr\"odinger equation by our recently developed Potential harmonic expansion method. We have earlier applied it successfully to 
study different properties of BEC~\cite{Pankaj1, Sudip, Pankaj2, Anindya, Anindya1, sudipPRA13, sudipEPJD13} and atomic clusters~\cite{Pankaj, TKD, Sudip1}. 
The methodology has already been described in detail in our earlier works~\cite{Tapan, Das, Kundu}. Hence here we describe it briefly for interested readers.

We consider a system of $N=(\mathcal{N}+1)$ Rb atoms, each of mass $m$ and interacting via two-body potential. 
The time-independent quantum many-body Schr\"odinger equation is given by
\begin{equation}
\Big[-\frac{\hbar^2}{2m}\sum_{i=1}^{N} \nabla_{i}^{2}
+\displaystyle{\sum_{i,j>i}^{N}} V(\vec{r}_{i}-\vec{r}_{j}) - E\Big]\Psi(\vec{r}_{1},...,\vec{r}_{N})=0 ,
\end{equation}
Where $E$ is the total energy of the system, 
$V(\vec{r}_{i}-\vec{r}_{j})$ is the two-body potential and 
$\vec{r}_{i}$ is the position vector of the $i$th particle. It is usual practice to decompose the motion of a many-body system into  
the motion of the center of mass and the relative motion of the particles in center of mass frame. In absence of any confinig potential 
the center of mass behaves as a free particle in laboratory frame and we set its energy as zero. Hence, after elimination of the center of mass motion and using standard 
Jacobi coordinates, defined as~\cite{Ballot, Fabre, MFabre}
\begin{equation}
\vec{\zeta}_{i}=\sqrt{\frac{2i}{i+1}}(\vec{r}_{i+1}-
\frac{1}{i}\sum_{j=1}^{i} \vec{r}_j) \hspace*{.5cm}
 (i=1,\cdots,{\mathcal N}),
\end{equation}
we obtain the equation for the relative motion of the atoms
\begin{equation}
\Big[-\frac{\hbar^{2}}{m}\sum_{i=1}^{\mathcal{N}} \nabla_{\zeta_{i}}^{2} + V_{int}(\vec{\zeta}_{1}, ..., \vec{\zeta}_{\mathcal{N}}) 
- E\Big]\Psi(\vec{\zeta}_{1}, ..., \vec{\zeta}_{\mathcal{N}}) =  0\hspace*{.1cm}, 
\end{equation} 
$V_{int}$ is the sum of all pair-wise interactions. Now it is to be noted that Hyperspherical harmonic expansion method 
is an {\it ab-initio} tool to solve the many-body Schr\"odinger equation where the total wave function is expanded in the complete 
set of hyperspherical basis~\cite{Ballot}. Although Hyperspherical harmonic expansion method is a complete many-body approach and includes all possible correlations, 
it is highly restricted to $N=3$ only. But for a diffuse cluster like Rb-cluster, only two-body correlation and pairwise interaction are important. 
Therefore we can decompose the total wave function $\Psi$ into two-body Faddeev component 
for the interacting $(ij)$ pair as 
\begin{equation}
\Psi=\sum_{i,j>i}^{N}\phi_{ij}(\vec{r}_{ij},r)\hspace*{.1cm}\cdot
\end{equation}
It is important to note that $\phi_{ij}$ is a function of two-body 
separation ($\vec{r}_{ij}$) only and the global 
hyperradius $r$, which is defined as $r = \sqrt{\sum_{i=1}^{N}\zeta_{i}^{2}}$. Thus the effect of two-body correlation comes 
through the two-body interaction in the expansion basis. $\phi_{ij}$ 
is symmetric under the exchange operator $P_{ij}$ for bosonic atoms and satisfy the 
Faddeev equation
\begin{equation}
\left[T-E_R\right]\phi_{ij}
=-V(\vec{r}_{ij})\sum_{kl>k}^{N}\phi_{kl}
\end{equation}
where $T$ is the total kinetic energy operator. In this approach, we assume that when ($ij$) pair interacts, the rest 
of the bosons are inert spectators. Thus the total hyperangular momentum 
quantum number as also the orbital angular momentum of the whole system 
is contributed by the interacting pair only. Next  the  $(ij)$th Faddeev 
component is expanded in the set of potential harmonics (PH) (which is 
a subset of  hyperspherical harmonic (HH) basis and sufficient for the expansion of $V(\vec {r}_{ij})$) 
appropriate for the ($ij$) partition as 
\begin{equation}
\phi_{ij}(\vec{r}_{ij},r)
=r^{-(\frac{3\mathcal{N}-1}{2})}\sum_{K}{\mathcal P}_{2K+l}^{lm}
(\Omega_{\mathcal{N}}^{ij})u_{K}^{l}(r) \hspace*{.1cm}\cdot
\end{equation}
$\Omega_{\mathcal{N}}^{ij}$ denotes the full set of hyperangles in the $3\mathcal{N}$-dimensional 
space corresponding to the $(ij)$  interacting pair and 
${\mathcal P}_{2K+l}^{lm}(\Omega_{\mathcal{N}}^{ij})$ is called the PH. It 
has an analytic expression:
\begin{equation}
{\mathcal P}_{2K+l}^{l,m} (\Omega_{\mathcal{N}}^{(ij)}) =
Y_{lm}(\omega_{ij})\hspace*{.1cm} 
^{(\mathcal{N})}P_{2K+l}^{l,0}(\phi) {\mathcal Y}_{0}(D-3) ;\hspace*{.5cm}D=3\mathcal{N} ,
\end{equation}
${\mathcal Y}_{0}(D-3)$ is the HH of order zero in 
the $(3\mathcal{N}-3)$ dimensional space spanned by $\{\vec{\zeta}_{1}, ...,
\vec{\zeta}_{\mathcal{N}-1}\}$ Jacobi vectors; $\phi$ is the hyperangle between the ${\mathcal{N}}$-th Jacobi vector 
$\vec{\zeta}_{\mathcal{N}}=\vec{r}_{ij}$ and the hyperradius $r$ and is given by
$\zeta_{\mathcal{N}}$ = $r\hspace*{0.1cm} \cos\phi$. For the remaining $(\mathcal{N}-1)$
noninteracting bosons we define hyperradius as
\begin{eqnarray}
 \rho_{ij}& = &\sqrt{\sum_{K=1}^{\mathcal{N}-1}\zeta_{K}^{2}}\nonumber\\
          &= &r \sin\phi \hspace*{.01 cm}\cdot
\end{eqnarray}
such that $r^2=r_{ij}^2+\rho_{ij}^2$. The set of $(3\mathcal{N}-1)$ quantum 
numbers of HH is now reduced to {\it only} $3$ as for the $(\mathcal{N}-1)$ 
non-interacting pair
\begin{eqnarray}
l_{1} = l_{2} = ...=l_{\mathcal{N}-1}=0,   & \\
m_{1} = m_{2}=...=m_{\mathcal{N}-1}=0,  &   \\
n_{2} = n_{3}=...n_{\mathcal{N}-1} = 0, & 
\end{eqnarray}
and for the interacting pair $l_{\mathcal{N}} = l$, $m_{\mathcal{N}} = m$ and  $n_{\mathcal{N}} = K$.
Thus the $3\mathcal{N}$ dimensional Schr\"odinger equation reduces effectively
to a four dimensional equation with the relevant set of quantum 
numbers: Energy $E$, orbital angular momentum quantum number $l$,
azimuthal quantum number $m$ and grand orbital quantum number $2K+l$
for any $N$. Substituting in Eq(4) and projecting on a particular PH, a set of 
coupled differential equation for the partial wave $u_{K}^{l}(r)$
is obtained
\begin{equation}
\begin{array}{cl}
\Big[-\frac{\hbar^{2}}{m} \frac{d^{2}}{dr^{2}} + \frac{\hbar^{2}}{mr^{2}}
\{ {\cal L}({\cal L}+1) 
+ 4K(K+\alpha+\beta+1)\} &\\
-E_R\Big]U_{Kl}(r) +\displaystyle{\sum_{K^{\prime}}}f_{Kl}V_{KK^{\prime}}(r)
f_{K^{\prime}l}
U_{K^{\prime}l}(r) = 0&\\
\hspace*{.1cm},
\end{array}
\label{eq.cde}
\end{equation}\\
where ${\mathcal L}=l+\frac{3N-6}{2}$, $U_{Kl}=f_{Kl}u_{K}^{l}(r)$, 
$\alpha=\frac{3N-8}{2}$ and $\beta=l+1/2$.\\
$f_{Kl}$ is a constant and represents the overlap of the PH for
interacting partition with the sum of PHs corresponding  to all 
partitions~\cite{MFabre}.
The potential matrix element $V_{KK^{\prime}}(r)$ is given by
\begin{equation}
V_{KK^{\prime}}(r) =  
\int P_{2K+l}^{lm^*}(\Omega_{\mathcal{N}}^{ij}) 
V\left(r_{ij}\right)
P_{2K^{\prime}+1}^{lm}(\Omega_{\mathcal{N}}^{ij}) d\Omega_{\mathcal{N}}^{ij} 
\hspace*{.1cm}\cdot
\end{equation}

Here we would like to point out that we did not require the additional short-range correlation function $\eta(r_{ij})$ for 
Rb clusters as was necessary for dilute BEC. A BEC is designed to be very dilute and hence confined by a harmonic oscillator potential of 
low frequency ($\sim 100$ Hz). The average interatomic separation is thus very large ($\sim 20000 a_0$) compared with the range of 
atom-atom interaction ($\sim 100 a_0$). Moreover the kinetic energy of the atoms is extremely small. Hence the effective interaction 
for large $r_{ij}$ is controlled by the $s$-wave scattering length ($a_s$)~\cite{Pethick}. This is achieved by the inclusion of the correlation 
function~\cite{Das, Kundu}. On the other hand, diffuse van der Waals clusters are weakly bound by the actual interatomic van der Waals 
potential (of range $\sim 10 a_0$), without any confinement. Hence no correlation function is needed. The average inter-particle separation is 
large enough, so that only two-body correlations are expected to be adequate,
 at least for light clusters.

\section{Results}
\subsection{Choice of interaction and calculation of many body effective potential}

\begin{figure}
  \begin{center}
    \begin{tabular}{cc}
       \resizebox{80mm}{!}{\includegraphics[angle=0]{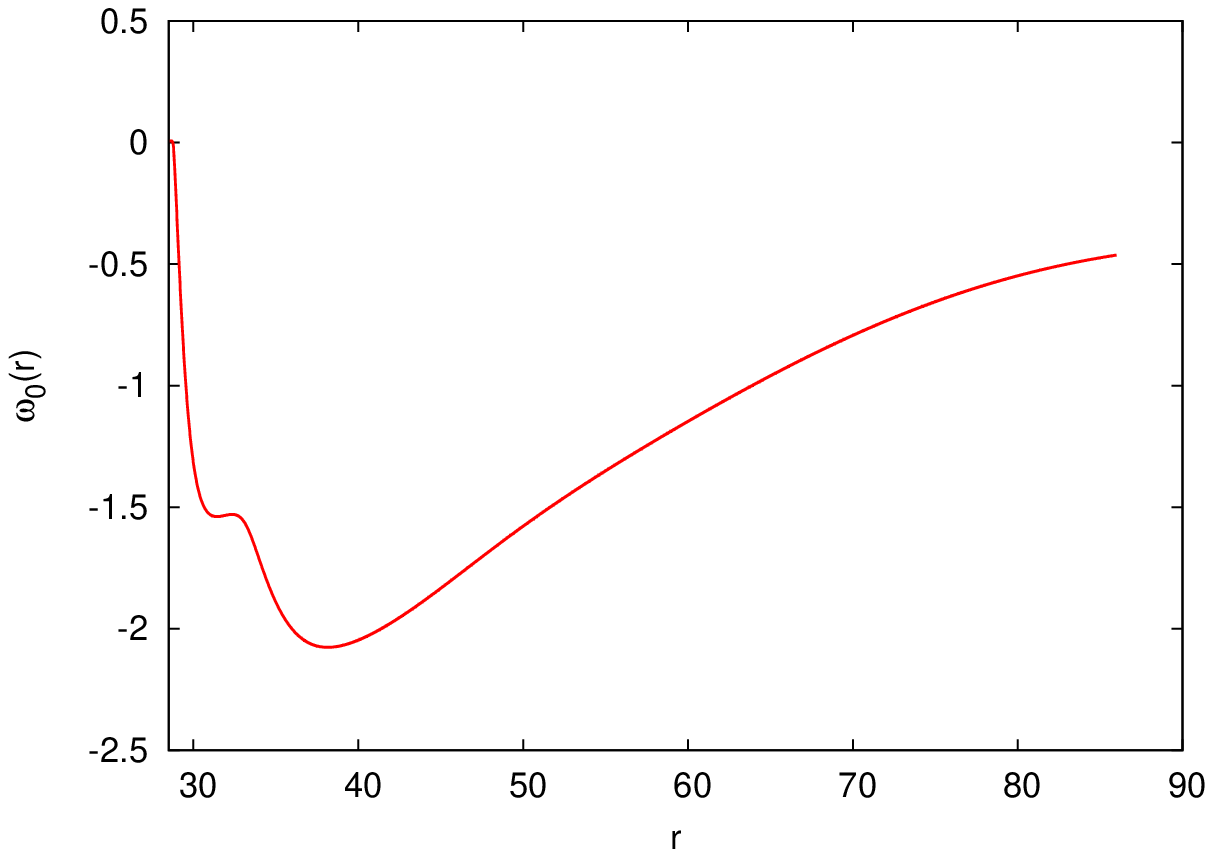}} & \\
         (a) $N$ =3 & \\
	\resizebox{80mm}{!}{\includegraphics[angle=0]{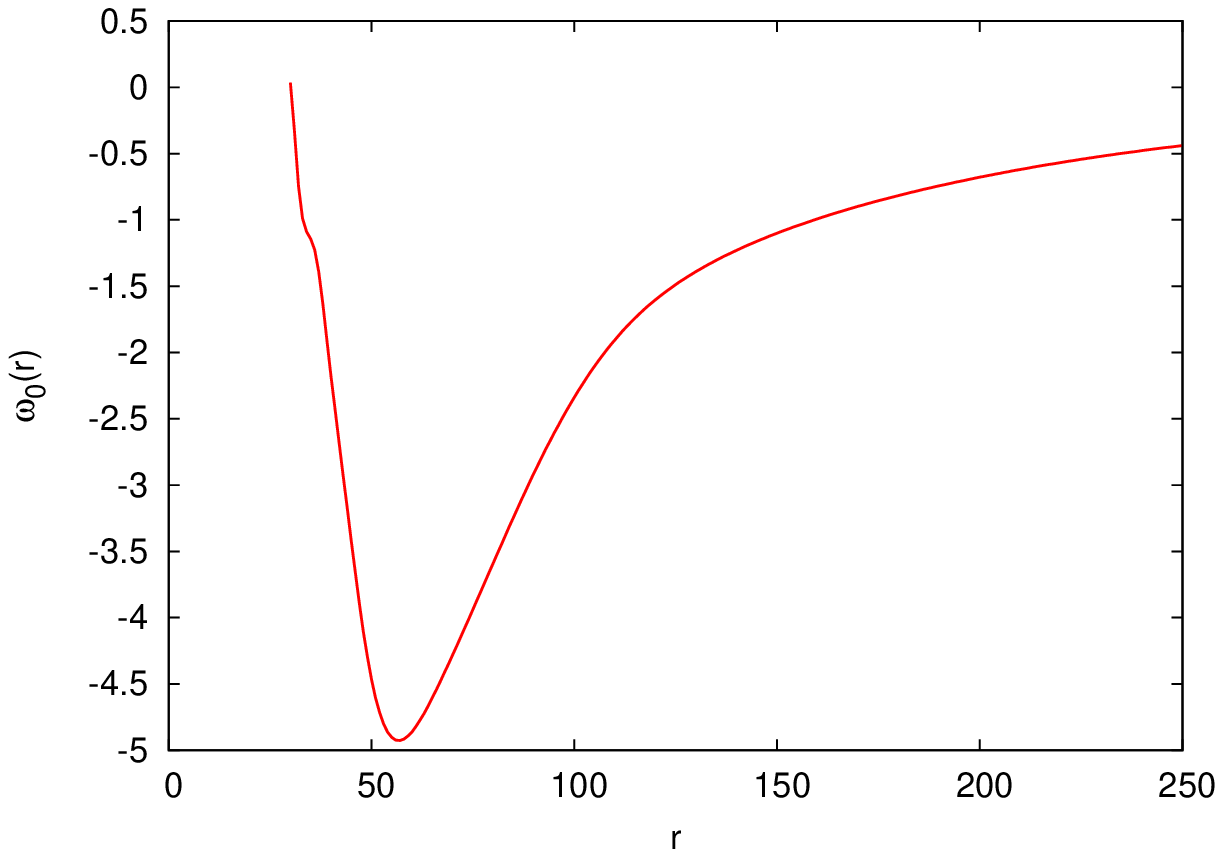}} &\\  
	 (b) $N$ =5 & \\
         \resizebox{80mm}{!}{\includegraphics[angle=0]{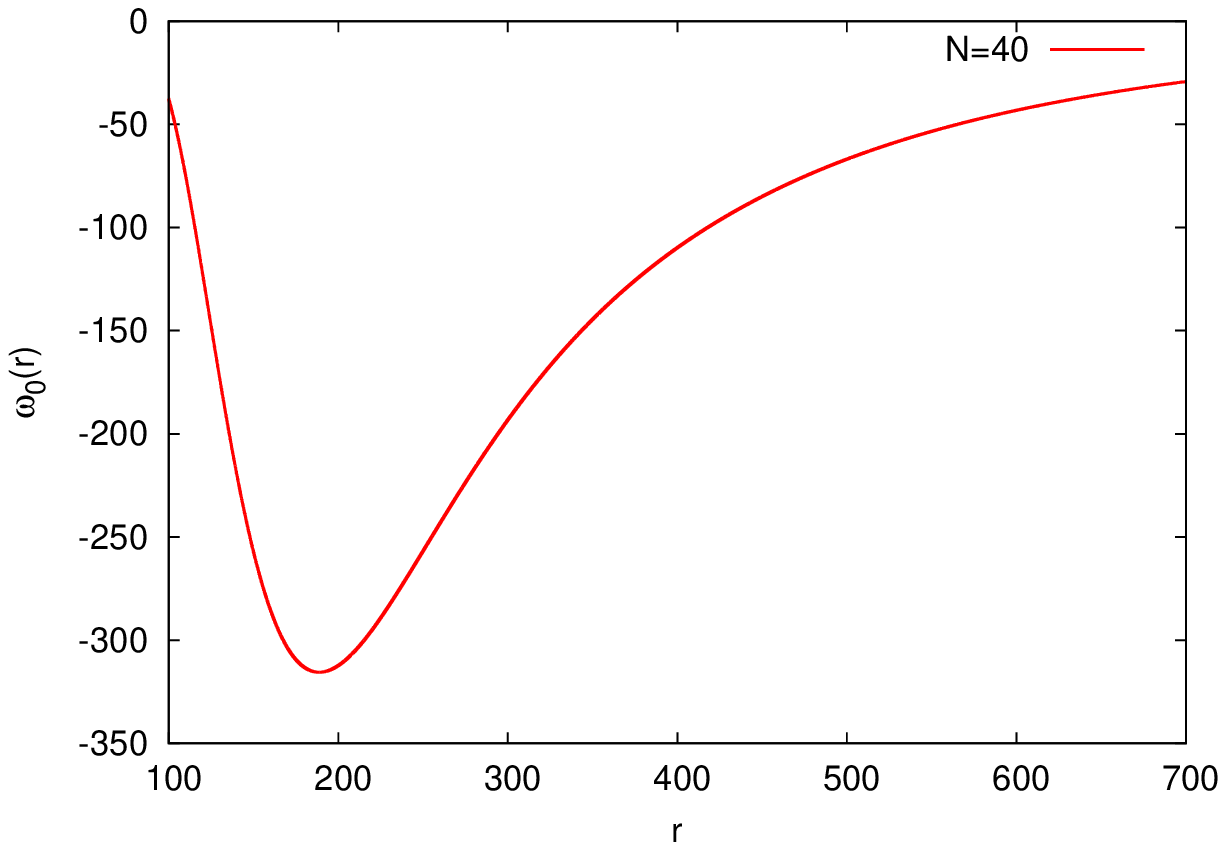}} & \\
          (c) $N$ = 40 & \\
 \end{tabular}
  \end{center}
\caption{(color online): Plot of the effective potential $\omega_0(r)$ for different cluster sizes, {\it viz.} $N=3$ [Panel (a)], 
$N=5$ [Panel (b)] and $N=40$ [Panel (c)].}
\label{fig.pot}
\end{figure}

As pointed earlier we choose the van der Waals potential with a hard core of radius $r_c$ as the interaction potential,
$V(r_{ij})$= $\infty$  for  $r_{ij} \leq r_c$ and
= $-\frac{C_6}{r_{ij}^6}$  for  $r_{ij}>r_c$. 
For Rb atoms, the value of $C_6$ is 2803 eV $\AA^6$~\cite{Pethick}. The unmanipulated scattering length corresponding to Rb-dimer 
is $a_s=100$ $a_0$. We obtain $a_s$ by solving the two-body Schr\"odinger equation for zero-energy~\cite{Kundu}. We adjust the hard core radius 
in the two-body equation to obtain the dimer scattering length. In the Fig.~1 of Ref.~\cite{Kundu} , we see the value of $a_s$ changes from negative 
to positive passing through an infinite discontinuity as $r_c$ decreases. Each discontinuity corresponds to one extra two-body bound state. 
We observe that tiny change in $r_c$ across the infinite discontinuity causes $a_s$ to jump from very large positive value to very large negative value. 
For our present calculation, we tune $r_c$ such that it corresponds to single bound state of the dimer. 
Thus calculated $r_c$ is $15.18 \AA$ for dimer scattering length of Rb atoms. With this set of values of $C_6$ and $r_c$, 
we next solve the coupled differential equation [\ref{eq.cde}] by hyperspherical adiabatic approximation~\cite{Coelho}. 
In hyperspherical adiabatic approximation, the hyperradial motion is assumed slow compared to hyperangular motion. For the hyperangular motion for a fixed value of $r$, 
we diagonalize the potential matrix together with the hypercentrifugal term. Thus the effective potential for the hyperradial motion is obtained as 
a parametric function of $r$. For the ground state of the system we choose the lowest eigenpotential $\omega_0(r)$ [corresponding eigen column vector 
being $\chi_{K0}(r)$] as the effective potential. We plot the effective potential $\omega_0(r)$ as a function of hyperradius $r$, at the dimer scattering 
length and for various cluster size $N=$3, 5 and 40 in Fig.~\ref{fig.pot}.
With increase in cluster size the depth of the eigen potential increases sharply which indicates stronger binding of the cluster. 
The average size of the cluster also increases with increases in $N$. The energy of the cluster is finally  obtained by solving the adiabatically 
separated hyperradial equation in the extreme adiabatic approximation
\begin{equation}
\left[-\frac{\hbar^{2}}{m}\frac{d^{2}}{dr^{2}}+\omega_{0}(r)-E_{R}
\right]\zeta_{0}(r)=0\hspace*{.1cm},
\end{equation}
subject to appropriate boundary condition.

In our earlier published work we have reported ground state and few low-lying excitation of Rb cluster with maximum size of $N=40$.~\cite{Pankaj}. 
However in the calculation of level statistics and spectral correlation we also need higher multipolar excitations. 
In our many-body picture the collective 
motion of the cluster is described by the effective potential. The excited states in this potential are denoted by $E_{nl}$ which 
corresponds to $n$th radial 
excitation with $l$th surface mode. Thus $E_{00}$ corresponds to the ground state and $E_{n0}$ are the different excitations for $l=0$. 
To calculate the higher 
levels with $l \neq 0$ we follow the next procedure. We have noted that for $l \neq 0$, a large inaccuracy is involved in the calculation 
of the off-diagonal 
potential matrix. As the main contribution to the potential matrix comes from the diagonal hypercentrifugal term we disregard the contribution 
coming from 
off-diagonal part. Thus we get the effective potential $\omega_l(r)$ for $l \neq 0$. 
Substituting $\omega_l(r)$ in Eq.~(14) we solve for different radial modes and repeat the numerical procedure for various $l$ to obtain the 
higher multipolar excitations. 

Before discussing the statistical behavior of the energy spectrum we should discuss how accurate our calculated energy levels are. 
It is to be noted that the potential
harmonic expansion method has been successfully applied in the calculation of collective excitations and thermodynamic properties of trapped bosons~\cite{Satadal}. 
For the investigation of thermodynamic properties we need to calculate a large number of energy levels. 
The calculated critical temperature and the condensate fraction 
are in good agreement with the experimental results~\cite{Satadal}. The effect of two-body correlations on thermodynamic properties of trapped bosons is 
also observed~\cite{Satadal}. Very recently we have also studied the energetics of diffuse $^{87}$Rb clusters~\cite{Pankaj} and also compared with the well studied He, 
Ne, and Ar clusters. Thus the calculated energy levels are accurate for further analysis. We also check for the convergence such that the error is 
considerably smaller than the mean level spacings. 

\subsection{Level-spacing statistics for different cluster sizes}

NNSD or $P(s)$ distribution is the most common observable which is used to study the short range fluctuation. 
Now to compare the statistical property of different parts of the spectrum we need to unfold them. By unfolding, the smooth part of the level 
density is removed, it basically maps the energy levels to another with the mean level density equal to $1$. For our present calculation we use 
polynomial unfolding of sixth order. We observe that for small cluster size with $N=3$ and $N=5$, as the effective potential is very shallow, 
the number of energy levels are very small and not sufficient for the calculation of NNSD. Instead, we also calculate the many-body collective levels 
including higher order excitations with different $l$. We then unfold each spectrum separately for a specific value of $l$ and then form an ensemble having 
the same symmetry. From the unfolded spectrum we calculate the nearest neighbour spacing $s$ as $E_{i+1}-E_i$ and calculate $P(s)$. $P(s)$ is 
defined as the probablity density of finding a distance $s$ between two adjacent levels. Uncorrelated spectra obey the Poisson 
statistics which gives exponential
distribution $P(s)=e^{-s}$. Whereas for system with time-reversal symmetry, level repulsion leads to the Wigner-Dyson distribution 
$P(s)= \frac{\pi}{2}s e^{\frac{-\pi s^2}{4}}$~\cite{Giannoni}.  

\begin{figure}
  \begin{center}
    \begin{tabular}{cc}
       \resizebox{80mm}{!}{\includegraphics[angle=0]{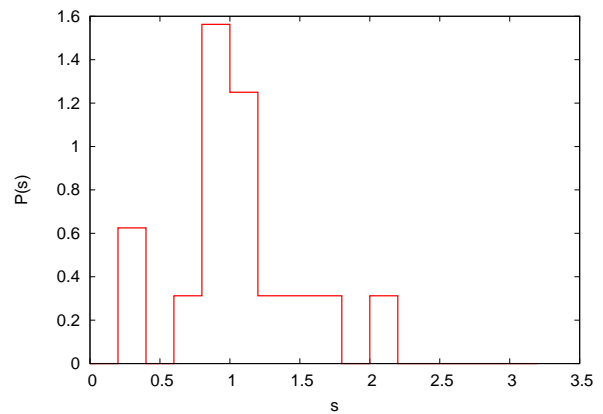}} &\\
       (a) lowest 22 levels & \\
	\resizebox{80mm}{!}{\includegraphics[angle=0]{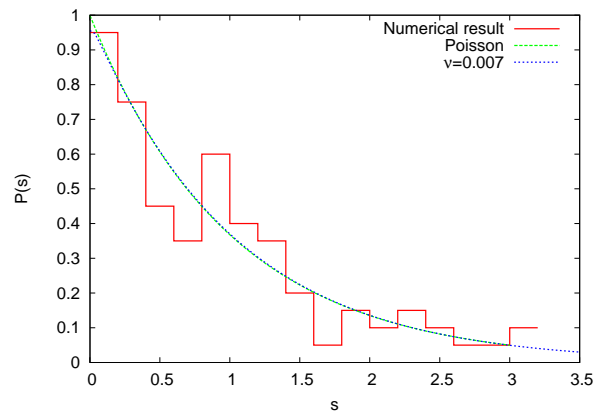}} &\\  
	(b) 300 $<$ levels $<$ 400 &\\
 \end{tabular}
  \end{center}
\caption{Plot of $P(s)$ distribution of lower [panel (a)] and higher [panel (b)] part of the spectrum of diffuse $^{87}$Rb cluster for $N=3$.
The green dashed curve in panel (b) represents the Poisson distribution whereas the blue dotted curve corresponds to the 
Brody distribution with the Brody parameter being $\nu=0.007$.}
\label{fig.ps_N=3}
\end{figure}

The $P(s)$ distribution of the unfolded spectrum with cluster size $N=3$ is plotted in Fig.~\ref{fig.ps_N=3}. 
We observe that $P(s)=0$ for very small $s$ and also for large $s$. In our earlier calculation of $^{87}$Rb diffuse cluster, 
we have calculated the several low-energy  
excitations. We have observed that due to the heavier mass of Rb atom, kinetic energy $<T>$ of Rb$_{N}$ clusters is small while the 
interaction energy $<V>$ is large. It 
implies that although the system is tightly bound, it is less correlated for smaller $N$. Thus  unlike the trapped bosons, the smaller 
diffuse cluster does not 
exhibit any degeneracy in the calculation of low-lying excitations. It is reflected in Fig.~\ref{fig.ps_N=3}(a) where we observe that 
$P(s)=0$ for very small $s$. 
The level spacing distribution for higher levels is shown in Fig.~\ref{fig.ps_N=3}(b) which indicates that for such small cluster, the energy levels 
are completely uncorrelated. Though it looks very similar to the Poisson distribution the peak value at $s=0$
is less than $1$. To determine how closely the histogram matches with the Poisson distribution we fit it with Brody distribution~\cite{Brody}
\begin{equation}
 P(\nu,s) = (1+\nu) a s^{\nu} \exp (-a s^{1+\nu})
 \label{eq.brody}
\end{equation}
where $a=[\Gamma(\frac{2+\nu}{1+\nu})]^{1+\nu}$ and $\nu$ is the Brody parameter. Depending on the value of the Brody parameter $\nu$, 
this distribution 
interpolates between the Poisson distribution $(\nu =0)$ and the Wigner distribution ($\nu=1$). Here we found $\nu = 0.007$. 
This implies that there is negligible 
correlation between the energy levels and the system is very close to regular. Actually for $N=3$ there are only $3$ interacting pair and the net 
attractive interaction is  very weak.

\begin{figure}[hbpt]
\vspace{-10pt}
\centerline{
\hspace{-3.3mm}
\rotatebox{0}{\epsfxsize=8cm\epsfbox{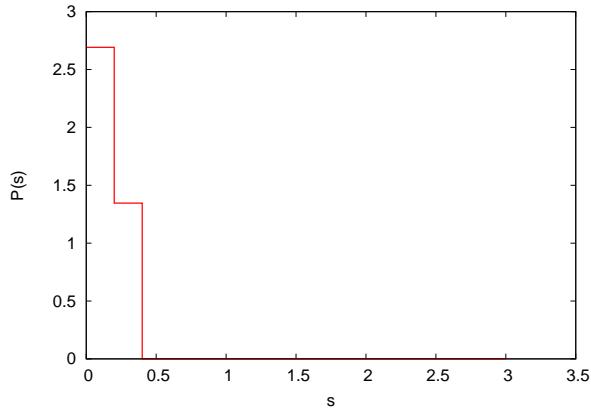}}}
\caption{Plot of the $P(s)$ distribution of lowest 30 levels of $^{87}$Rb cluster for $N=5$.}
\label{fig.ps_N=5_lowest30}
\end{figure}

Next to study the effect of inter-atomic interaction we gradually increase the effective interaction. 
We can vary the effective interaction either by tuning the 
scattering length $a_s$ or by changing the number of bosons. Here we increase the number of bosons to $N=5$. 
It is already known from the earlier study of $^{4}$He 
cluster that $\Delta E = E_{N+1}-E_N$ decreases smoothly as a function of $N$ which indicates the saturation in 
the density and predicts liquid-drop behavior in 
$^4$He cluster with larger $N$~\cite{Pankaj}. However diffuse Rb cluster which is the system of our present interest is 
dilute and less compact which indicates sharp change in $\Delta E$ with change in cluster size. 
The average size of the cluster also increases. Thus the cluster with $N=5$ is more tightly bound, stable and more 
correlated compared with 
the cluster size $N=3$. Due to more correlation in the energy spectrum, we can expect the very closely spaced energy levels which 
leads to the quasi-degeneracy.
This is reflected in Fig.~\ref{fig.ps_N=5_lowest30} where we plot the $P(s)$ distribution for the lowest 30 levels. 
The sharp peak in the first bin near $s=0$ 
clearly exhibits the signarure of quasidegeneracy. This peak is known as Shnirelman peak~\cite{Shnirelman}.
\begin{figure}
  \begin{center}
    \begin{tabular}{cc}
       \resizebox{80mm}{!}{\includegraphics[angle=0]{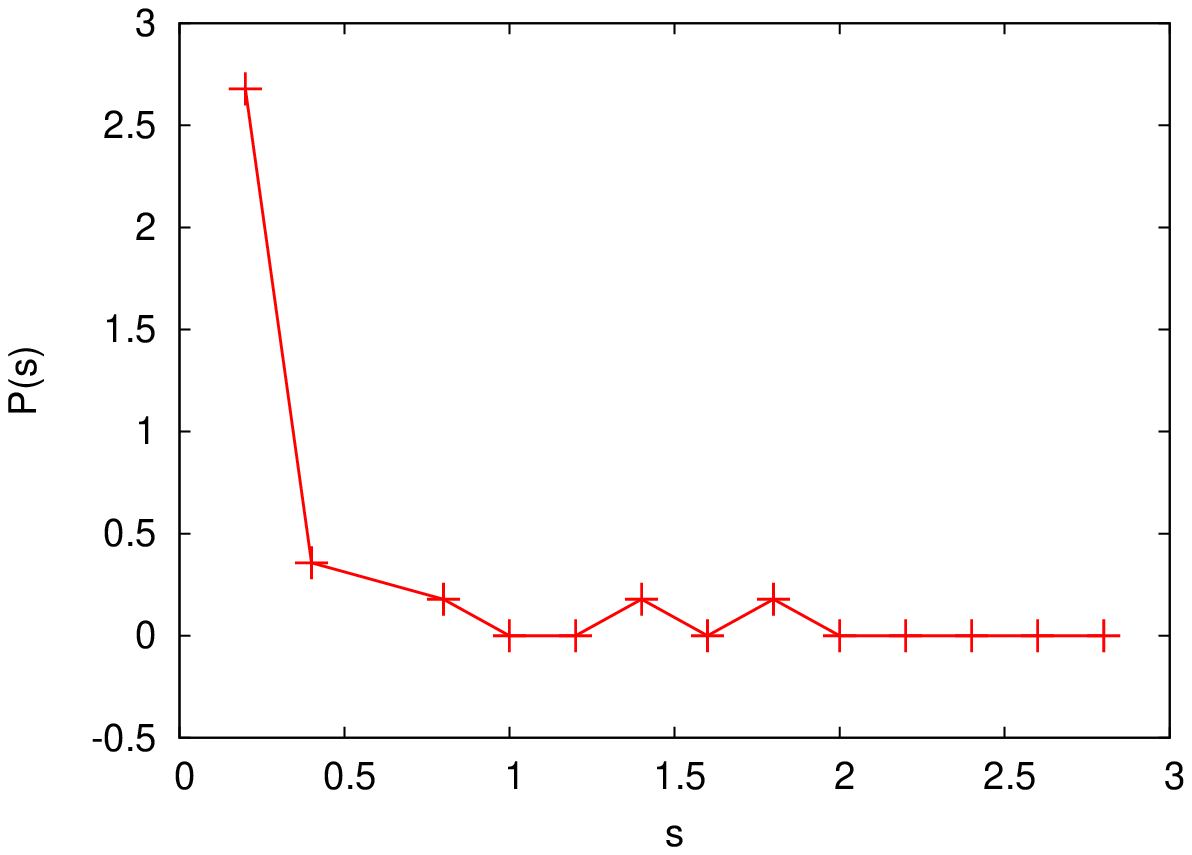}} & \\
       (a) & \\
       \resizebox{80mm}{!}{\includegraphics[angle=0]{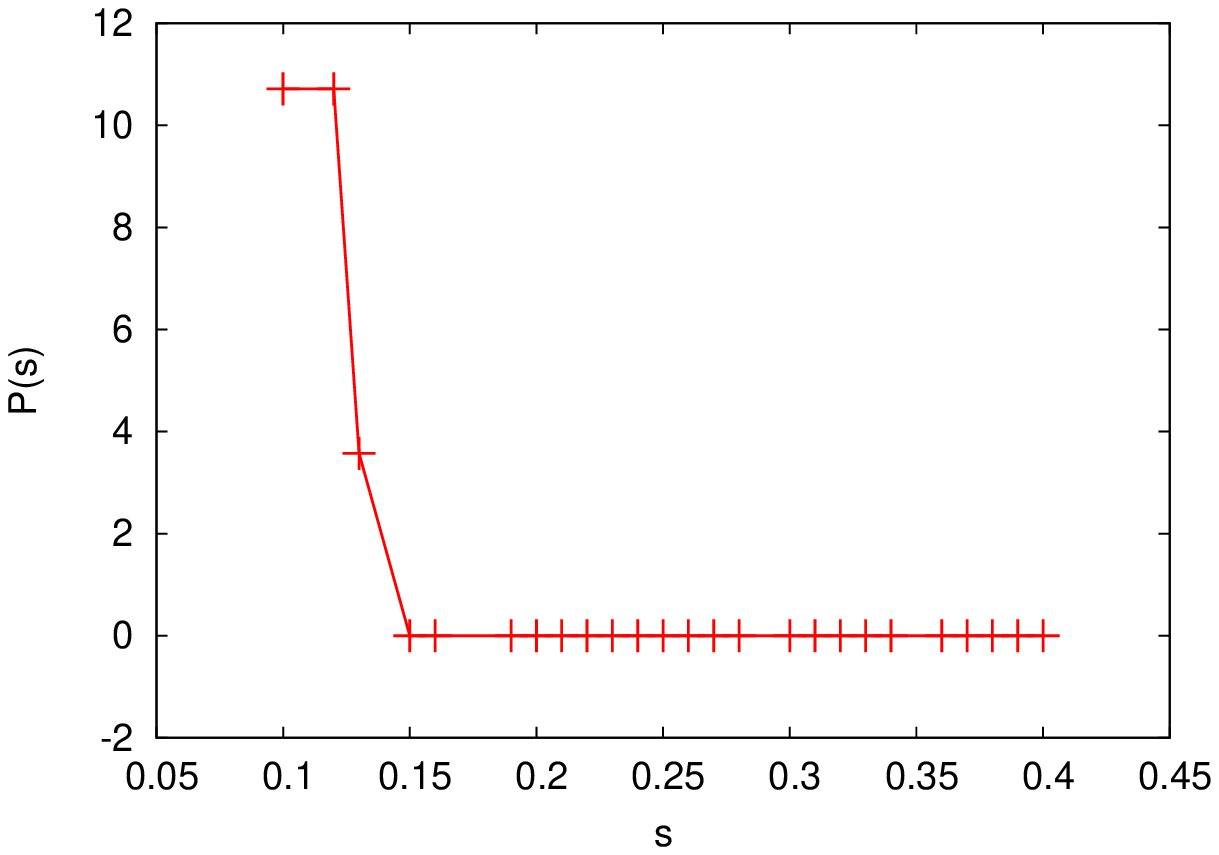}} & \\  
	(b)  & \\
 \end{tabular}
  \end{center}

 \caption{The Structure of Shnirelman peak observed for lowest $30$ levels of $^{87}$Rb cluster with $N=5$ is shown in finer detail. 
The bin size in panel(a) is 0.2. In panel (b) only the peak is further zoomed by taking the bin size 0.01.}
 \label{fig.Shnirelman}
\end{figure}

\begin{figure}
\vspace{-10pt}
\centerline{
\hspace{-3.3mm}
\rotatebox{0}{\epsfxsize=8cm\epsfbox{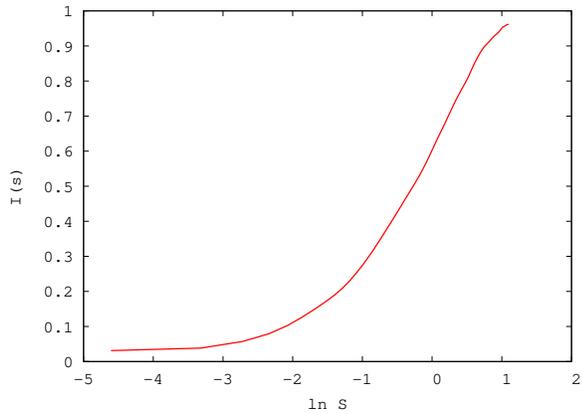}}}
\caption{Plot of the integral level spacing distribution $I(s)$ vs $\ln s$ for $N=5$.}
\label{fig.Is}
\end{figure}

For better understanding of the structure of Shnirelman peak, we plot the same histogram 
in Fig.~\ref{fig.Shnirelman} as in Fig.~\ref{fig.ps_N=5_lowest30} in finer details. Reducing the bin size gradually, a huge peak appears in the first 
bin which demonstrates the existence of global quasidegeneracy. The peak has a finite width which is further associated with Poisson tail. 
The resolution of the peak is further studied as the integral level spacing distribution $I(s) = N P(s)$ (here $N$ being the number of levels), 
normalized to unity. We plot $I(s)$ as a function of $\ln s$ in Fig.~\ref{fig.Is}. The linear dependence 
between $I$ and $\ln s$ is shown in the left most part of Fig.~\ref{fig.Is} which represents the structure of the Shnirelman peak. 
Whereas the rightmost steep increase of $I(s)$ corresponds to the Poisson tail. 
\begin{figure}
  \begin{center}
    \begin{tabular}{cc}
      \resizebox{80mm}{!}{\includegraphics[angle=0]{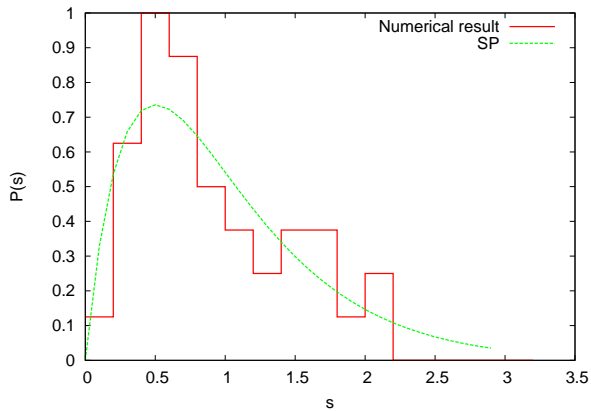}} & \\
	(a) $40 <$ level $< 80$ & \\
      \resizebox{80mm}{!}{\includegraphics[angle=0]{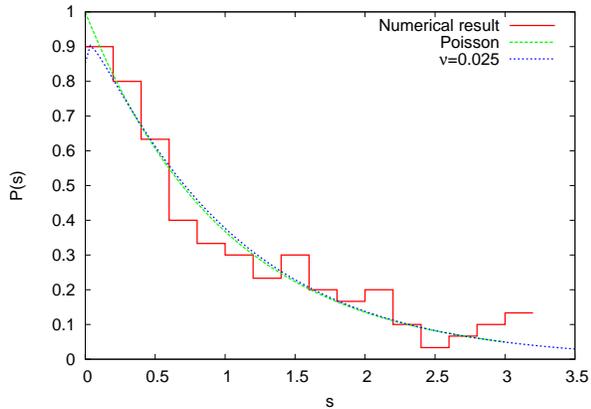}} & \\
	(b) $850 <$ level$ <1000$ & \\ 
  \end{tabular}
  \end{center}

 \caption{Plot of the $P(s)$ distribution for middle and higher levels for $N=5$. The green dashed curve in panel (a) represents the 
semi-Poisson distribution and that in panel (b) presents the Poisson distribution. The blue dotted curve in panel (b) corresponds to the 
Brody distribution with the Brody parameter $\nu =0.025$.}
\label{fig.ps_N=5}
\end{figure}
However for the higher levels we observe that the system exhibits pseudo-integrability. It is reflected in Fig.~\ref{fig.ps_N=5}(a) 
where we observe the 
semi-Poisson (SP) distribution. For comparison, in the same 
figure we plot the analytic expression of SP statistics given by $P(s) = 4 s e^{-2s}$~\cite{Antonio}. We observe the level repulsion at smaller 
values of $s$ ($s << 1$), where $P(s) \propto s$ and asymptotic decay of $P(s)$ is exponential. The SP distribution is observed within a narrow 
intermediate region between the quasi-degenerate regime and the completely integrable regime. $P(s)$ distribution for the higher levels are plotted 
in the Fig.~\ref{fig.ps_N=5}(b) which is again very similar to Poisson distribution. We again fit the Brody distribution with the histogram and find the 
Brody parameter $\nu =0.025$. The observation of SP distribution and increase in the value of Brody parameter $\nu$ clearly manifests 
the enhanced effect of 
inter-atomic correlation with increase in cluster size. However we fail to give any physical reason which causes this SP and Poisson statistics. 
As pointed earlier, for smaller cluster size only $l=0$ effective potential is not enough to calculate sufficient number of levels for the 
study of $P(s)$ distribution. So the findings of SP statistics may be physically acceptable whose origin is not clear to us or it may be due to overlap of 
several $l$ values.
\begin{figure}[hbpt]
\vspace{-10pt}
\centerline{
\hspace{-3.3mm}
\rotatebox{0}{\epsfxsize=8cm\epsfbox{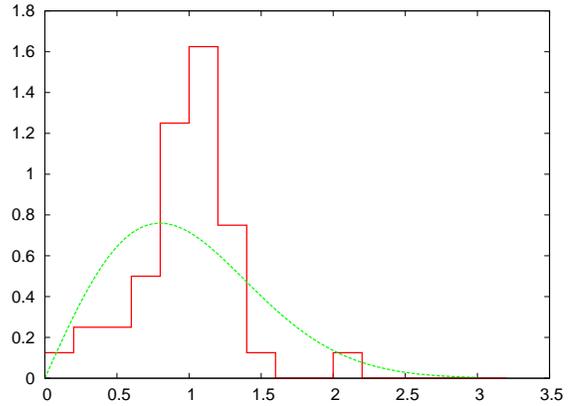}}}
 \caption{Plot of the nearest neighbour level spacing distribution $P(s)$ of the higher portion (160-200 levels) of the spectrum of diffuse 
van der Waals cluster for $N=40$. The red smooth histogram represents our numerical result and the green dashed curve represents the Wigner distribution. }
\label{fig.ps_N=40}
\end{figure}
Thus to get further insight we significantly change the cluster size to $N=40$ where only $l=0$ effective potential is deep enough to 
support sufficient number of states for calculation of $P(s)$ distribution. We plot the $P(s)$ distribution in Fig.~\ref{fig.ps_N=40}. We observe similarity with 
Wigner distribution as very small value of $P(s)$ near $s=0$ signifies the level repulsion.
However the peak at $s=1$ overshoots $1$. The large peak at $s=1$ signifies large accumulation of levels with level spacing $s=1$. Though we tried 
to fit the histogram again with Brody distribution we fail 
to appropriately fit it. Now it is worthy to mention that Guhr and Weidenm\"uller~\cite{Guhr} proposed a modified uniform spectrum 
in terms of a deformed GOE, which combines uniform, GOE and Poisson. As the $P(s)$ distribution of Fig.~\ref{fig.ps_N=40} 
is quite similar to Fig.~1, Fig.~2, and Fig.~3 of Ref.~\cite{Guhr}, the use of deformed GOE may be an ideal step for future investigation. 
As the Fig.~\ref{fig.ps_N=40} does not match with the Wigner 
distribution we conclude that the Hamiltonian is not chaotic. However the deformed GOE type distribution signifies the system is strictly nonintegrable and exhibits 
strong interatomic correlation. Thus it is indeed required to calculate the energy level correlation which we discuss in the following section.

\subsection{Energy level correlation}

\begin{figure}
  \begin{center}
    \begin{tabular}{cc}
      \resizebox{80mm}{!}{\includegraphics[angle=0]{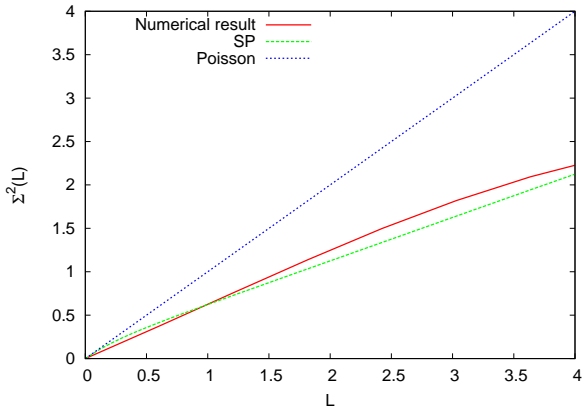}} & \\
	(a) $N=5$, $40 < level < 80$& \\
      \resizebox{80mm}{!}{\includegraphics[angle=0]{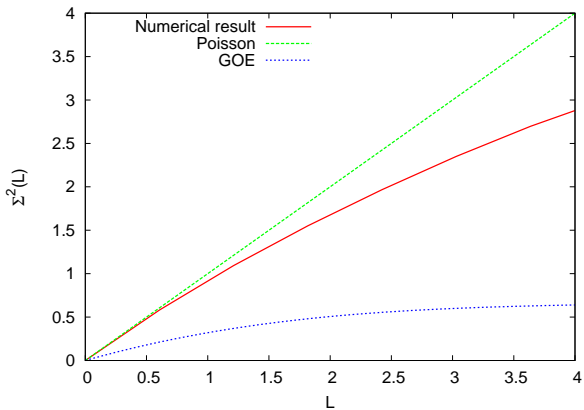}} & \\
	(b) $N=5$, $850 < level < 1000$  & \\ 
     \end{tabular}
   \end{center}
 \caption{(color online) Plot $\Sigma^2(L)$ vs $L$ for intermediate and higher part of the spectrum for $N=5$.}
 \label{fig.sigma2_N=5}
\end{figure}

\begin{figure}[hbpt]
\vspace{-10pt}
\centerline{
\hspace{-3.3mm}
\rotatebox{0}{\epsfxsize=8cm\epsfbox{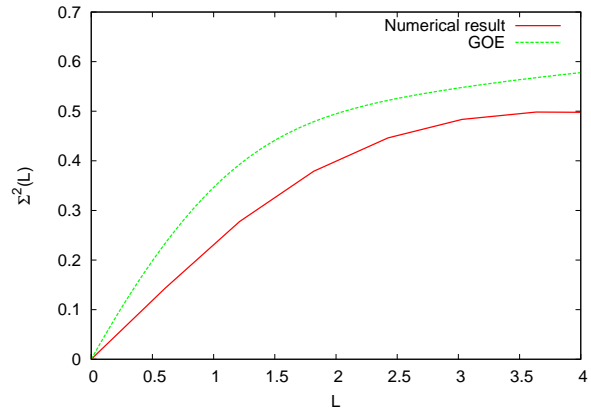}}}
 \caption{ (color online) Plot $\Sigma^2(L)$ vs $L$ for the spectrum of Rb cluster with $N=40$.}
 \label{fig.sigma2_N=40}
\end{figure}

So far we have considered only the NNSD which is commonly used to characterize the short-range fluctuations in the spectrum. 
However in order to confirm our findings 
of the effect of correlation on the spectral properties and to investigate how the correlation gradually builds in with the increase in cluster size
which makes the system too complex, we study the long range correlations of the spectrum. The level number variance $\Sigma^2(L)$ is 
the most commonly used observable to characterize correlations between pair of levels. It mainly determines the long-range fluctuations in the spectrum. 
It is defined as the average variance of the number of levels in the energy interval containing an average number of $L$ levels and is calculated as
\begin{equation}
 \Sigma^2(L) = <(N(E+L)-N(E)-L)^2>
\end{equation}
where $<>$ represents the average over the energy value $E$ and $N(E)$ determines the number of eigen energy levels below $E$. For the uncorrelated 
Poisson statistics $\Sigma^2(L) =L$, whereas for GOE, $\Sigma^2(L)$ increases logarithmically with $L$. From the earlier study of level spacing 
distribution it has been observed that for $N=3$ the system exhibits features which are very close to the non-interacting limit. 
We have also observed the Poisson distribution in the level statistics of higher levels. However the most interesting observation is the 
semi-Poisson distribution for the intermediate part of the spectrum for $N=5$. The corresponding $\Sigma^2(L)$ is plotted in Fig.~\ref{fig.sigma2_N=5}(a). 
It approximately increases linearly as $L/2$ which is the value of number variance $\Sigma^2(L)$ of SP distribution. 
Then we plot $\Sigma^2(L)$ for higher part of the spectrum in Fig.~\ref{fig.sigma2_N=5}(b). It  is approximately proportinal to $L$ indicating that 
the system is correlated but does not exhibit any level repulsion. This further confirms the findings of the 
Poisson distribution in the $P(s)$ distribution. For strongly correlated cluster with $N=40$ we observe that $\Sigma^2(L)$ approximately 
increases logarithmically with $L$ [Fig.~\ref{fig.sigma2_N=40}]. 
This feature is close to GOE results. However there are significant differences between our  numerical results and the Wigner surmise. It again indicates 
that the system does not show full chaos though it exhibits strong non-integrability.

\begin{figure}[hbpt]
 \vspace{-10pt}
\centerline{
\hspace{-3.3mm}
\rotatebox{0}{\epsfxsize=8cm\epsfbox{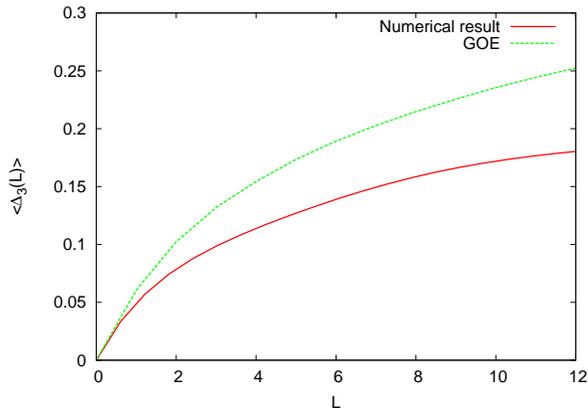}}}
 \caption{ (color online) Plot $<\Delta_3(L)>$ vs $L$ for the spectrum of Rb cluster with $N=40$.}
 \label{fig.delta3_N=40}
\end{figure}

\begin{figure}
  \begin{center}
    \begin{tabular}{cc}
      \resizebox{80mm}{!}{\includegraphics[angle=0]{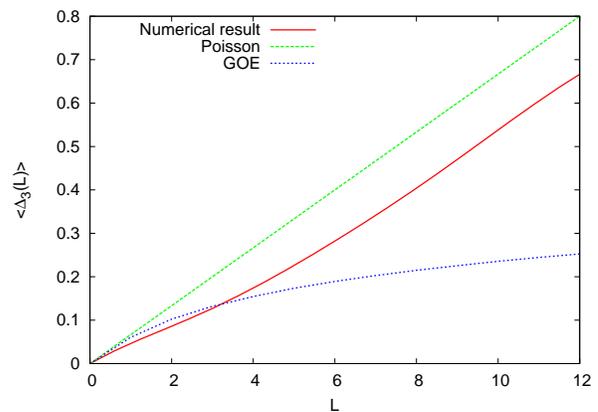}} & \\
	 (a) $40 < lelvel <80$ & \\
      \resizebox{80mm}{!}{\includegraphics[angle=0]{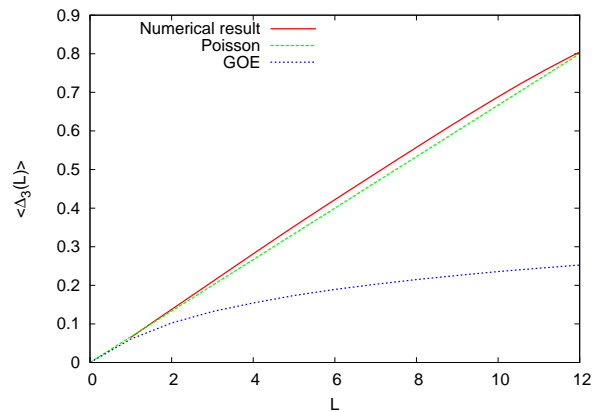}} & \\
	(b) $850 < level < 1000$ & \\
     \end{tabular}
\end{center}
 \caption{(color online) Plot of $<\Delta_3(L)>$ vs $L$ for intermediate [panel (a)] and higher part [panel(b)] of spectrum for $N=5$.}
 \label{fig.delta3_N=5}
\end{figure}

The other important observable to characterise long-range correlation is $\Delta_3-$statistics~\cite{Mehta}. Given an energy interval 
$[\alpha, \alpha+L]$ of length $L$, it is defined as the least square deviation of the staircase function $\hat{N}(E_i)$ from the best straight line fitting it:
\begin{equation}
\Delta_3(\alpha;L)=\frac{1}{L}Min_{A,B}\int_{\alpha}^{\alpha+L}[\hat{N}(E_i)-AE_i-B]^2 dE_i
\end{equation}
It is customary to use the average values of $\Delta_3(L)$. Thus $\Delta_3-$statistics, averaged over energy intervals, measures 
the deviation of the unfolded spectrum from the equidistant spectrum and hence it gives information on  
the rigidity of spectrum or spectral stiffness. For uncorrelated Poisson spectra $<\Delta_3(L)> \propto L$ whereas for Wigner spectra 
$<\Delta_3(L)> \propto \log L$. Our 
calculated numerical results for $N=40$ is shown in Fig.~\ref{fig.delta3_N=40}. Though it looks similar to GOE distribution, it is significantly lower than
the GOE results which confirms our earlier observation for large cluster. The approach of $<\Delta_3(L)>$ towards the GOE behavior is valid only upto 
$L \approx 2$. The similar kind of observation was made in Fig. 6 of Ref.~\cite{Guhr} where deformed GOE behavior is noted in $<\Delta_3(L)>$. 
It indicates that for large cluster size, 
the levels are strongly correlated. Whereas for smaller cluster ($N=5$), we observe that $<\Delta_3(L)>$ distribution gradually approaches to Poisson as 
we move upward in the spectrum. For a small
intermediate region of the spectrum $<\Delta_3(L)>$ lies between the GOE and Poisson distribution [Fig.~\ref{fig.delta3_N=5}(a)] whereas for the upper levels it almost 
perfectly follows the Poisson distribution [Fig.~\ref{fig.delta3_N=5}(b)] which indicates that the spectrum has turned soft. 

\subsection{ Quotients of successive spacings}

Before concluding the paper, we present in this Section,  as a test of the
observations made in Sections III B, the results of the analysis of the
distribution of quotients of successive level spacings  [denoted by $P(r)$], a
measure introduced recently, that is independent of the unfolding function and
the  unfolding procedure \cite{Huse2007,ABGR-2013}. Note that in all the
analysis presented in Sections III B and III C, we have employed a sixth-order
polynomial for the density of levels for unfolding. The P(r) distribution and
the related averages allow for a more transparent comparison with experimental
results than the traditional level spacing distribution and this measure is
particularly important for many-body systems as the theory for the eigenvalue
(level) densities for these systems is usually not available. In the recent
past, this measure was used in analyzing many-body localization
\cite{Huse2007,OPH2009,Pal-10,Iyer-12} and also in quantifying the distance from
integrability on finite size lattices \cite{koll2010,Coll2012}. More recently,
using $P(r)$ it is established conclusively that embedded random matrix 
ensembles for many-body systems, generated by random interactions in the
presence of a mean-field, follow GOE for strong enough two-body interaction
\cite{CV-2013}.

Given a ordered set of the energy levels $E_n$, the nearest neighbor spacing
$s_n= E_{n+1}- E_n$ and the probability distribution of the ratios
$r_n=s_n/s_{n-1}$ is $P(r)$ subject to normalization $\int P(r)dr=1$. 
If the system is in integrable domain (described by Poisson NNSD), then the $P(r)$ is given by 
\begin{equation} 
P_P(r)=\frac{1}{(1+r)^2}
\label{eq.prpoi} 
\end{equation}
and if the system is chaotic (described by GOE), then the $P(r)$ is given by
Wigner-like surmise \cite{ABGR-2013}, 
\begin{equation}
 P_W(r)= \frac{27}{8} \frac{r+r^2}{(1+r+r^2)^{5/2}}\;. 
\label{eq.prgoe}
\end{equation} 
The average value of r, i.e. $\left\langle r \right\rangle$, is $1.75$ for GOE and is $\infty$
for Poisson. It is also possible to consider $\tilde{r}_n = \frac{\min(s_n,
s_{n-1})}{\max(s_n, s_{n-1})} = \min(r_n, 1/r_n)$. The average value of
$\tilde{r}$, i.e. $\left\langle \tilde{r}\right\rangle$, is $0.536$ for GOE and $0.386$ for
Poisson.

\begin{figure}
  \begin{center}
    \begin{tabular}{cc}
       \resizebox{80mm}{!}{\includegraphics[angle=0]{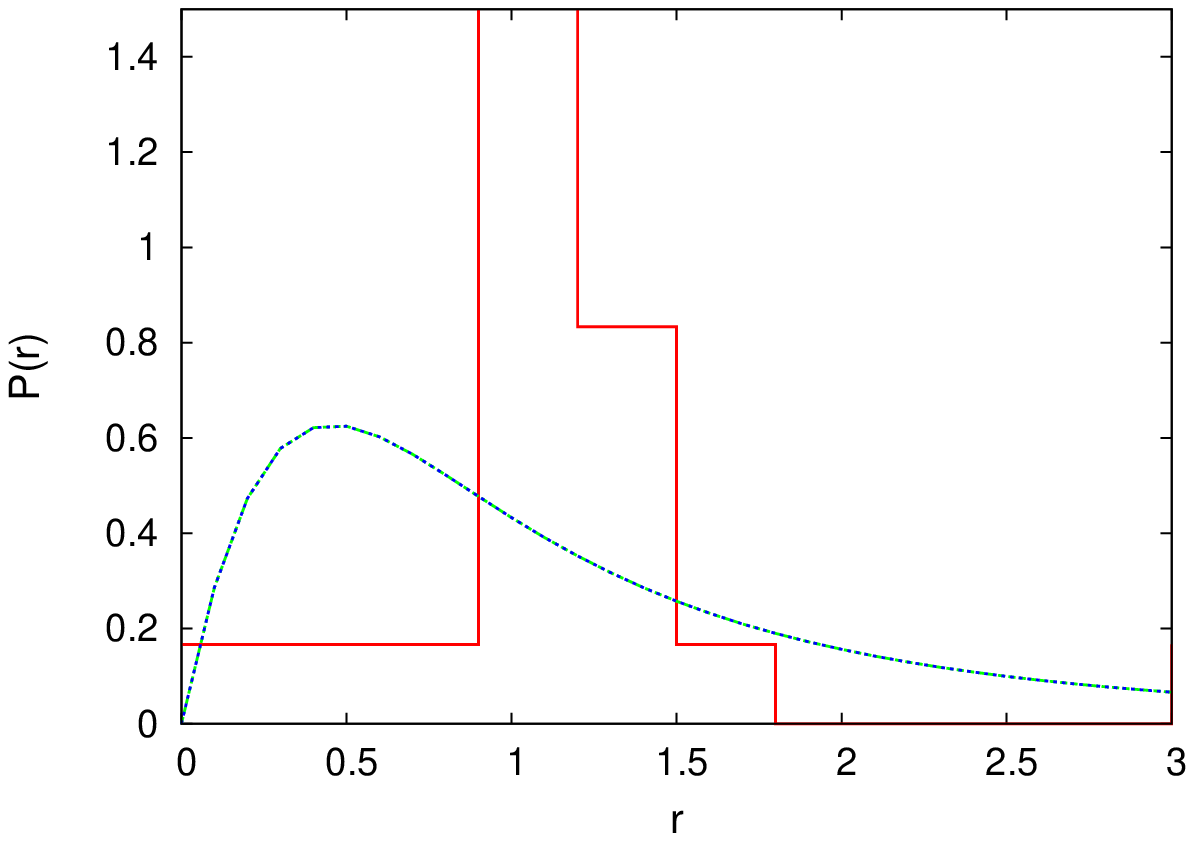}} & \\
         (a) $N$ =3 & \\
	\resizebox{80mm}{!}{\includegraphics[angle=0]{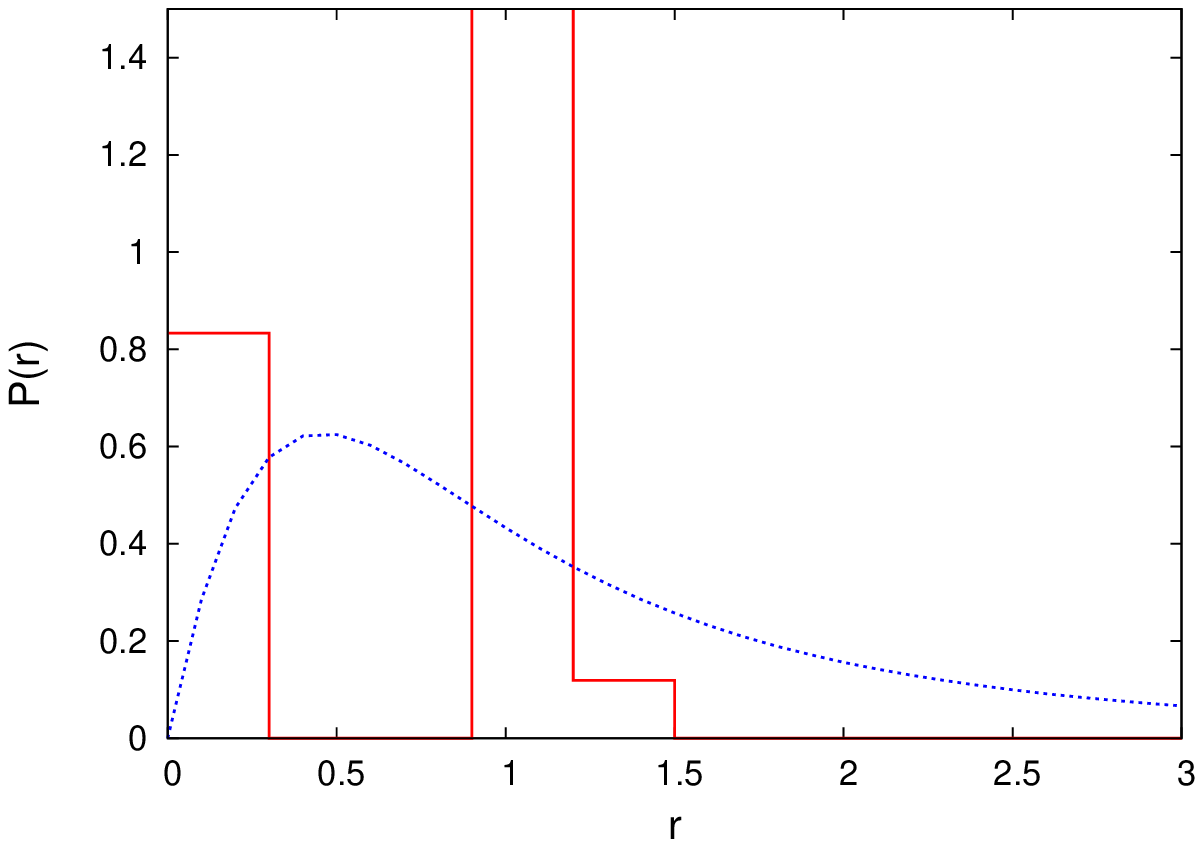}} &\\  
	 (b) $N$ =5 & \\
         \resizebox{80mm}{!}{\includegraphics[angle=0]{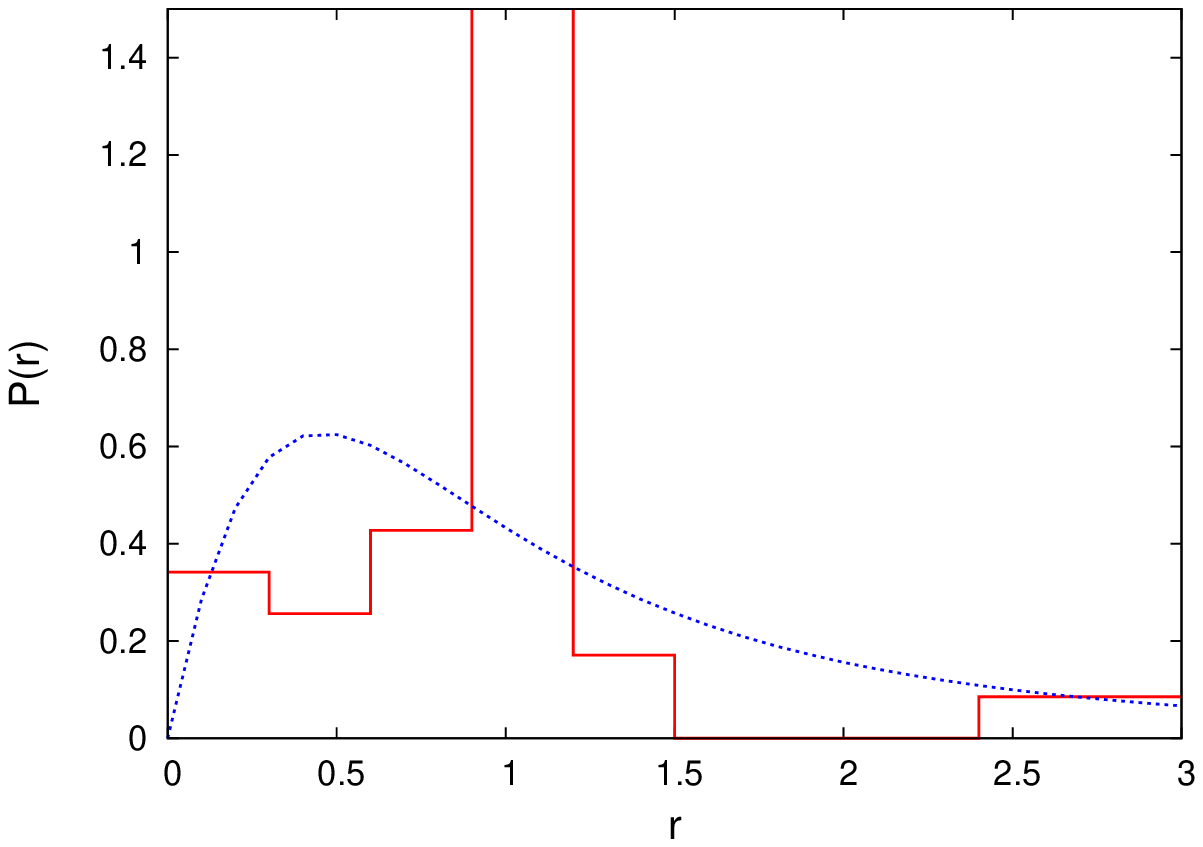}} & \\
          (c) $N$ = 40 & \\
 \end{tabular}
  \end{center}

\caption{Distribution of the ratio of consecutive level spacings $P(r)$ of the 
spectrum of diffuse $^{87}$Rb for cluster sizes (a) $N=3$ with lowest $22$ levels,
(b) $N=5$ with lowest $30$ levels  and (c) $N=5$ with levels $40-80$. Result for
GOE (blue curve) is also shown.}

\label{prdr1}
\end{figure}

\begin{figure}
  \begin{center}
    \begin{tabular}{cc}
       \resizebox{80mm}{!}{\includegraphics[angle=0]{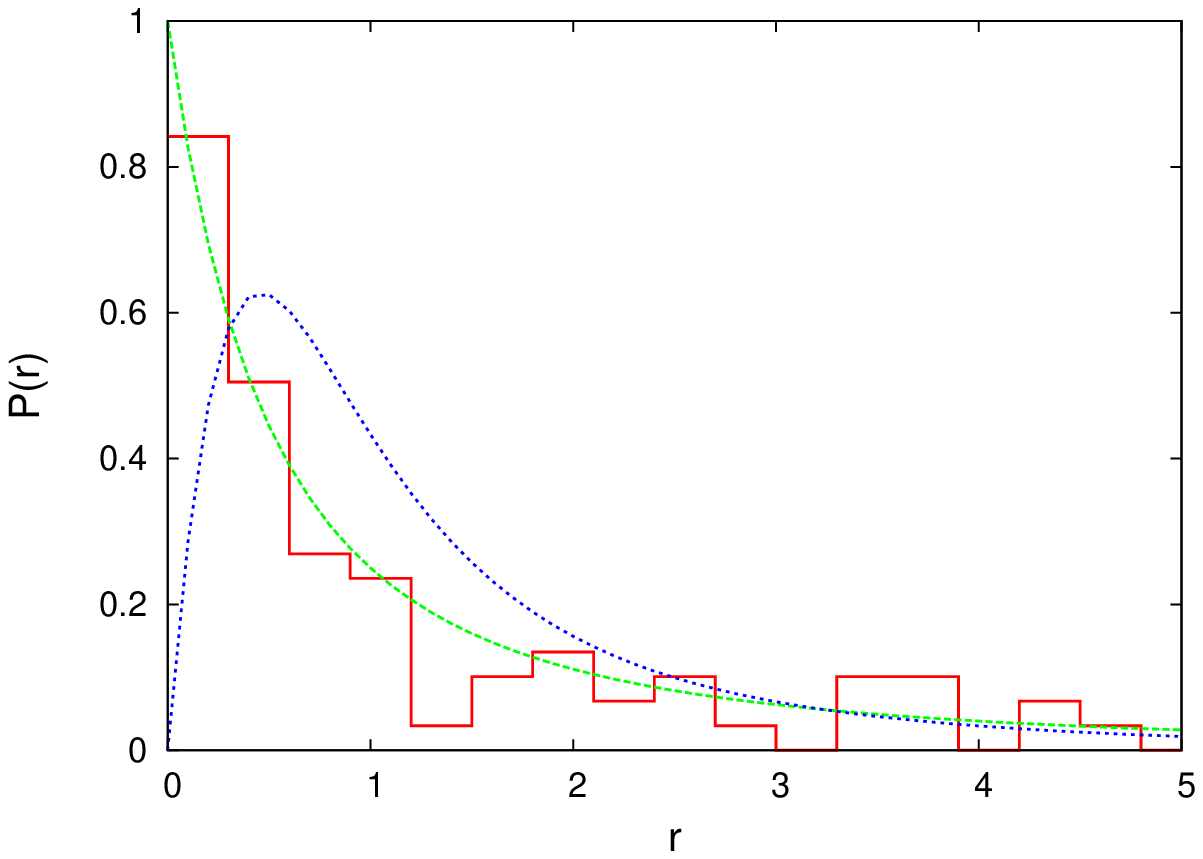}} & \\
         (a) $N$ =3 & \\
	\resizebox{80mm}{!}{\includegraphics[angle=0]{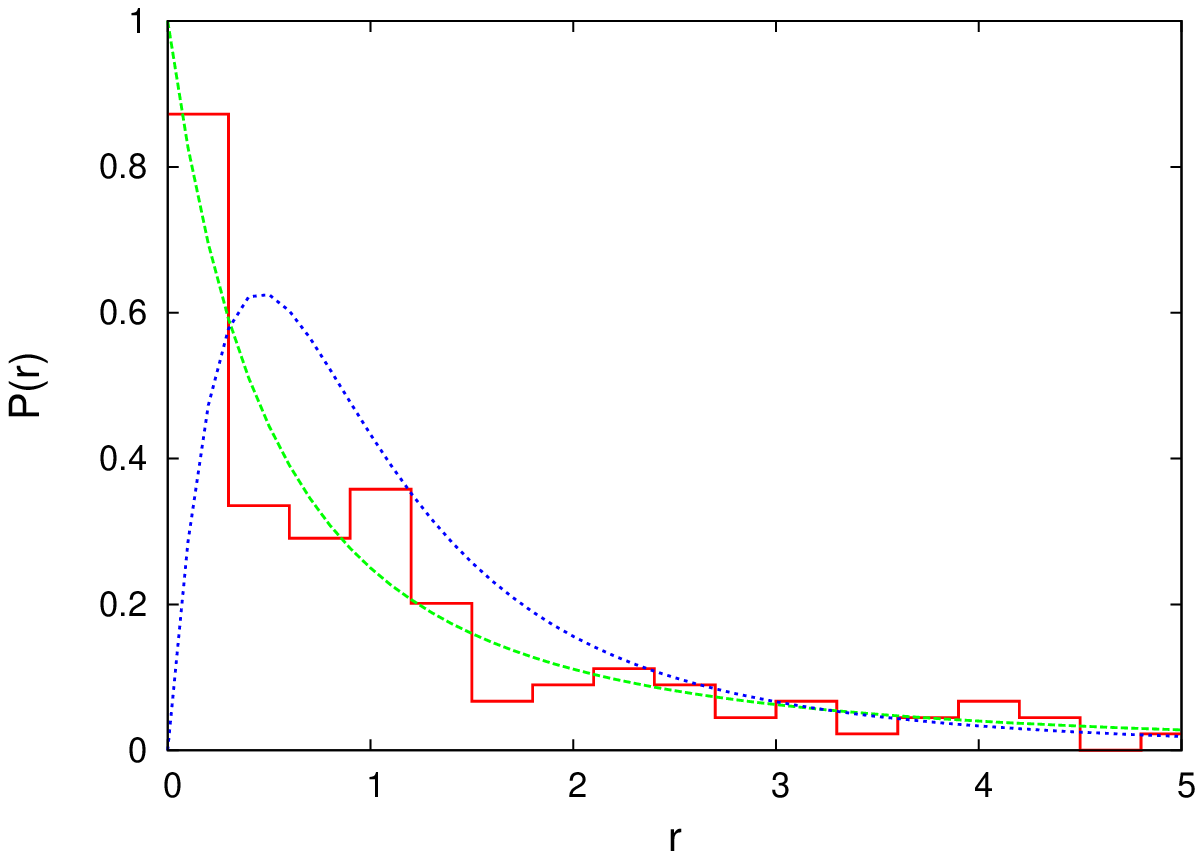}} &\\  
	 (b) $N$ =5 & \\
         \resizebox{80mm}{!}{\includegraphics[angle=0]{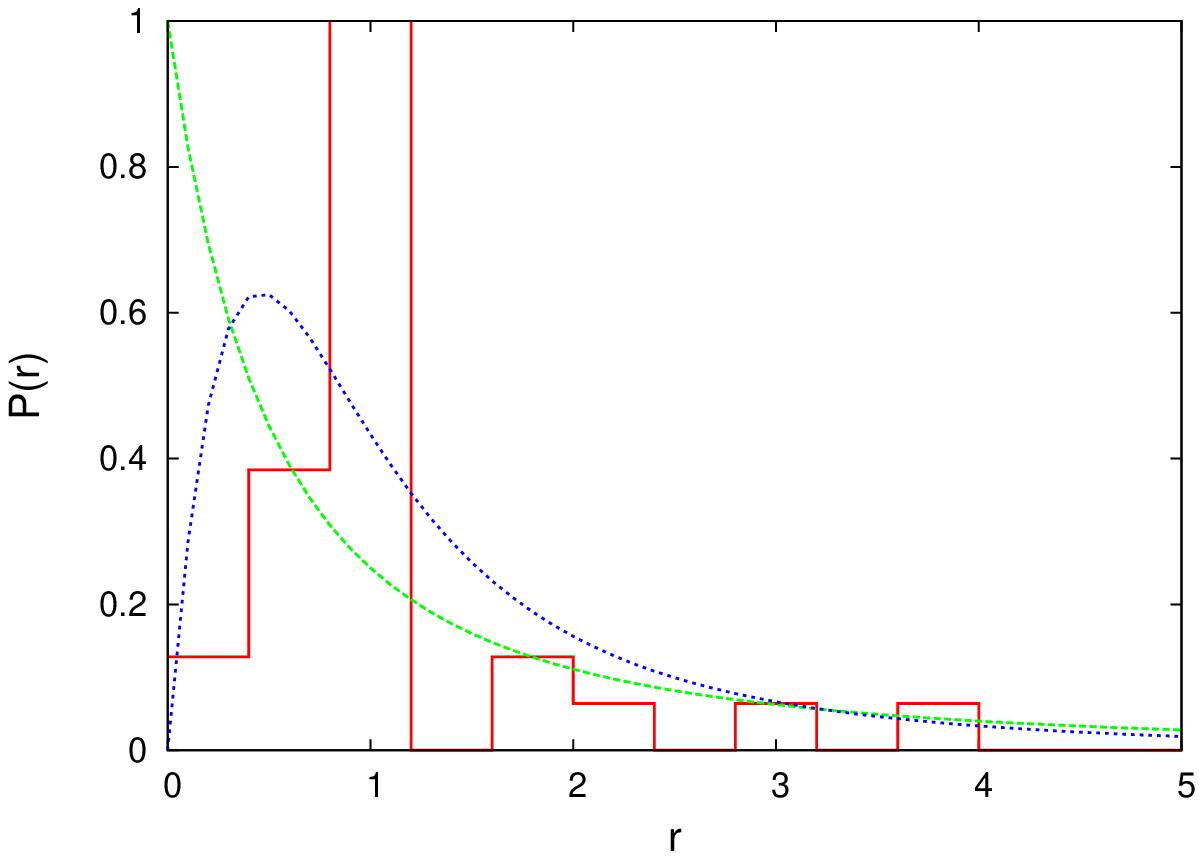}} & \\
          (c) $N$ = 40 & \\
 \end{tabular}
  \end{center}

\caption{Distribution of the ratio of consecutive level spacings $P(r)$ of the 
spectrum of diffuse $^{87}$Rb for cluster sizes (a) $N=3$ with levels 
$300-400$, (b) $N=5$ with levels $850-1000$ and (c) $N=40$ with levels
$160-200$.  Results for Poisson (green curve) and GOE (blue curve) are also
shown.}

\label{prdr2}
\end{figure}

Some results for $P(r)$ vs $r$ for the spectrum of diffuse $^{87}$Rb cluster with the same cluster sizes as above {\it viz.} 
$N$ = 3, 5 and 40, are shown in Figs. \ref{prdr1} and \ref{prdr2}. Moreover,  we have
also calculated the averages $\left\langle r \right\rangle$ and $\left\langle \tilde{r}\right\rangle$ and results
are given in Table 1. For $N=3$ with levels 1-22, there is a peak at $r \sim 1$
as seen from Fig. \ref{prdr1}a. Similarly for levels 300-400, $P(r)$ is close to
Poisson form as shown in Fig. \ref{prdr2}a. These results are consistent with
the NNSD results in Figs. 2a and 2b respectively. In addition, the results for
$\left\langle r\right\rangle$ and $\left\langle \tilde{r}\right\rangle$ given in Table 1 are also in agreement
with these observations. Turning to $N=5$, with levels 1-30 the $P(r)$ shows
peaks at $r \sim 0$ and $r \sim 1$ (see Fig. \ref{prdr1}b) and for  quantifying
this structure, it is necessary to derive $P(r)$ that corresponds to Shnirelman
peak. Going to levels 40-80, it is seen from Fig. \ref{prdr1}c that $P(r)$
exhibits level repulsion with $P(r) \sim 0$ for $r \sim 0$ but the form of
$P(r)$ shows clear deviations from the GOE result given by Eq. (\ref{eq.prgoe}).
In order to compare with the conclusion drawn from NNSD in Fig.~6a, it is
necessary to derive the formula for $P(r)$ for pseudo-integrable systems (these
systems give semi-Poisson form for NNSD). Turning to levels 850-1000, it is
clearly seen from Fig. \ref{prdr2}b that the $P(r)$ is close to Poisson and this
is in complete agreement with NNSD shown in Fig. 6b. Further, for $N=40$ the
$P(r)$ curve shows level repulsion and it is closer to GOE than to Poisson (see
Fig. \ref{prdr2}c). Also,  the values of $\left\langle r \right\rangle$ and $\left\langle \tilde{r}\right\rangle$
(shown in Table 1) are close to GOE results. Thus N=40 example exhibits level
repulsion as seen in the NNSD result. Combining all these observations, we
conclude that the results deduced from NNSD analysis are consistent with those
obtained from $P(r)$ analysis and thus the unfolding procedure used in
Sections III B and III C can be considered to be good.

\begin{table}[h]
\centering
\caption{Values of averages $\left\langle \tilde{r}\right\rangle$ and $\left\langle r \right\rangle$  for 
various cluster size $N$.}

\begin{tabular}{llcc}
\hline
\hline
& & $\left\langle \tilde{r}\right\rangle$ & $\left\langle r \right\rangle$ \\
N=3 & levels (1-22)  & 0.76 &1.168\\
&levels (300-400) & 0.34 &204151\\
& & & \\
N=5 & levels (1-30)  & 0.48 &6.18\\
&levels (40-80)  & 0.64 &1.64\\
&levels (850-1000) &0.39 &144078\\
& & &\\
N=40 &levels(160-200)  & 0.76 &1.43\\
\hline
& & &\\
GOE &&0.5359 &1.75\\
Poisson&& 0.3863&$\infty$\\
\hline
\end{tabular}
\label{table1}
\end{table}

\section{Conclusions}

Study of energy level statistics plays an important role in elucidating the universal properties of quantum systems. Berry and Tabor conjectured that 
the eigenenergy levels of a quantum system whose classical dynamics shows integrability, must exhibit the fluctuation property as determined by the uncorrelated 
Poisson statistics. This is in sharp contrast with the BGS conjecture which asserts that the fluctuation property of energy levels of a quantum system whose classical 
dynamics should exhibit GOE (or GUE or GSE) statistics. However, complicated quantum
many-body systems often lie between these two contrasting conjectures.

Thus the purpose of present paper is to consider a relatively complex quantum system whose experimental realization is possible. The van der Waals bosonic 
cluster is such a quantum system which starts to be more and more complex with increase in cluster size. The above mentioned contrasting conjectures have been examined 
thoroughly by using various statistical observables like NNSD, level number variance $\Sigma^2(L)$ and the spectral rigidity $\Delta_3(L)$. These observables 
highlight the short and long range correlation, level repulsion, level clustering and how the features of the above observables crucially depend on the cluster 
size are also focussed. Our detailed numerical analysis reveals that for smaller cluster 
when the system is very close to integrability, Berry and Tabor conjecture is followed. For 
large cluster although we observe similar to BGS conjecture, however deviation occurs. For large clusters the system becomes strongly correlated but does not exhibit 
true chaos. However our present study reveals that the deformed GOE type of distribution may be suitable for future investigation.
\hspace*{.5cm}
\begin{center}
{\large{\bf{Acknowledgements}}}
\end{center}
This work is supported by the Department of Atomic
Energy (DAE), government of India, through Grant No.
2009/37/23/BRNS/1903.
SKH acknowledges the Council of Scientific and Industrial Reaserch (CSIR), India for a senior research fellowship through NET (Grant No: 08/561(0001)/2010-EMR-1). 
NDC acknowledges financial support from the University Grants Commission (UGC), India [Grant No: F.40-425/2011 (SR)]. 
SKH also acknowledges hospitality of the Maharaja Sayajirao University of Baroda, Vadodara, India during a recent visit for this work.

\end{document}